\documentstyle[preprint,aps,epsfig]{revtex}

\newcommand{\beq}{\begin{eqnarray}}
\newcommand{\eeq}{\end{eqnarray}}
\newcommand{\beqs}{\begin{eqnarray}}
\newcommand{\eeqs}{\end{eqnarray}}
\newcommand{\bary}{\begin{array}}
\newcommand{\eary}{\end{array}}

\newcommand{\figpos}{p}      

\def\eq#1{{\rm eq.}~(\ref{#1})}

\def\Eq#1{{\rm Eq.}~(\ref{#1})}

\def\eps{\varepsilon}

\def\Re{{\cal R}e}

\def\Oc{{\cal O}}

\def\eV{{\rm~eV}}
\def\MeV{{\rm~MeV}}
\def\GeV{{\rm~GeV}}
\def\TeV{{\rm~TeV}}
\def\vR{v_R}

\def\npb#1{Nucl.\ Phys.\ {\bf B\,#1}}

\def\plb#1{Phys.\ Lett.\ {\bf B\,#1}}

\def\prd#1{Phys.\ Rev.\ {\bf D\,#1}}
\def\prl#1{Phys.\ Rev.\ Lett. {\bf#1}}

\def\epjc#1{Eur.~Phys.~J.\ {\bf C\,#1}}

\def\pha#1{{\tt hep-ph/#1}}
\def\phb#1{{[{\tt hep-ph/#1}]}}
\def\exa#1{{\tt hep-ex/#1}}
\def\exb#1{{[{\tt hep-ex/#1}]}}
\def\aspha#1{{\tt astro-ph/#1}}
\def\asphb#1{{[{\tt astro-ph/#1}]}}

\def\nucexb#1{{[{\tt astro-ph/#1}]}}

\def\gsim{\ \rlap{\raise 3pt \hbox{$>$}}{\lower 3pt \hbox{$\sim$}}\ }
\def\lsim{\ \rlap{\raise 3pt \hbox{$<$}}{\lower 3pt \hbox{$\sim$}}\ }

\def\putFig#1#2#3#4#5#6#7 
{
 \newpage
 \begin{figure}[\figpos]
 \LARGE
 $#7$
 \begin{center}
 \mbox{\epsfig{figure=#1.eps,angle=0,width=#3cm,height=#4cm}}
 $#6$
 \end{center}
 \vspace{1.5cm}
 \caption{#2}
 \label{#1}
 \end{figure}
}

\begin{document}

\preprint{\vbox{\hbox{WIS/17/01-AUG-DPP}
                \hbox{hep-ph/0108176}}}

\title{On Neutrino Masses and a Low Breaking Scale of Left-Right Symmetry} 

\author{Oleg Khasanov and Gilad Perez  \\ 
        \small \it Department of Particle Physics,
        Weizmann Institute of Science,
        Rehovot 76100, Israel}
\maketitle

\begin{abstract}
In left-right symmetric models (LRSM) the light neutrino masses arise 
from two sources:
the seesaw mechanism and a VEV of an SU(2)$_L$ triplet. If the
left-right symmetry breaking, $v_R$, is low, $v_R\lsim15\TeV$, 
the contributions to the light neutrino masses
from both the seesaw mechanism and 
the triplet Yukawa couplings are expected to be 
well above the experimental bounds.
We present a minimal LRSM with an additional
U(1) symmetry in which the masses induced by the two sources are below
the eV scale and the two-fold problem is solved.
We further show that, if the U(1) symmetry is also responsible for 
the lepton flavor structure, the model
yields a small mixing angle within the first two lepton generations. 
\end{abstract}

\vspace{1cm}

\section{Introduction}
Among the various new physics (NP) models that give neutrino masses, the 
left-right symmetric (LRS) framework is, in many senses, 
one of the most attractive. 
A left-right symmetric model (LRSM) is
based on the gauge group SU(2)${}_R \times\,$SU(2)$_L
\times\,$U(1)${}_{\rm B-L}$~\cite{LRSM}.
At the high energy scale of the LRS breaking, 
$v_R$, various new fields and interactions are present.
In many LRSM, {\it e.g.} LRSM embedded in GUT models~\cite{GUT},
the scale $v_R$ is much higher than the electroweak (EW)
breaking scale, $k$, and the low energy effective
Lagrangian is similar in many aspects to that of the Standard Model
(SM). In such cases
present and near future experiments will not be able to directly probe the
NP.
We therefore investigate whether a natural LRSM with $v_R\lsim 15\TeV$ 
can be constructed consistent 
with the experimental data on the lepton flavor parameters.
In particular we require the active neutrino masses to
be below the eV scale,
\beq
m_{\nu_i}\lsim 1\eV
\,,
\label{nu}
\eeq
with $i=1,2,3$. This is motivated by the cosmological~\cite{Cos}, 
and direct~\cite{PDG,Beta} upper bounds on the active neutrino masses 
and in particular by the ones deduced from the atmospheric and solar
experiments assuming hierarchical neutrino  
masses~\cite{SNO,Nuexp,Analysis}.
The charged lepton masses are~\cite{PDG},
\beq
m_e\simeq0.51~{\rm MeV}\,, \ \
m_\mu\simeq105.7~{\rm MeV}\,,\ \
m_\tau\simeq1777~{\rm MeV}\,.
\label{mcl}
\eeq

In section \ref{natural}
we discuss a two-fold problem related to 
neutrino masses which one faces when dealing with a relatively low scale of
$v_R$; 
In section \ref{toy} we present a minimal LRSM (MLRSM)
with an additional U(1) symmetry, which solves
the above problem. 
In section \ref{U1} we show that if the same U(1) symmetry accounts
also for the lepton flavor structure such a model  
cannot yield a large mixing angle (LMA) between 
the first two generations.
Comments and conclusions are given in section \ref{conclusion}.


\section{Naturalness of the MSLRM with a low $\vR$}\label{natural}

Below we investigate the consequences of having a low value for
the LR symmetry breaking scale, 
\beq
v_R\lsim 15\TeV
\label{vR}\,.
\eeq
The Lagrangian of the lepton sector in the MLRSM  is given by:
\beq
{\cal L}&=& {\cal L}_0+
f\bar L_L \phi_1 L_R+g \bar L_L \phi_2 L_R+
h(\Delta_L L_L L_L+\Delta_R L_R L_R)+V(\phi,\Delta)
+h.c.
\label{L}
\,,
\eeq
where ${\cal L}_0$ contains the kinetic and the gauge interaction
terms for the various fields. The field
$L$ is a lepton doublet, the $\phi$ field is a Higgs bi-doublet,
$\phi_1=\pmatrix{h^0_1&h_1^+\cr h_2^-&h^0_2}$, 
$\phi_2=\tau_2 \phi_1^* \tau_2$,
$\Delta_L$ [$\Delta_R$] is a triplet of the SU(2)$_L$ [SU(2)$_R$]
gauge group 
and the scalar potential, $V(\phi,\Delta)$, is given below.

In the general case the scalar fields, $\Delta_{L,R}$  and 
$\phi$, develop complex VEVs, 
\beq
 \langle \Delta_{L,R} \rangle &=& \left( 
                            \begin{array} {ccc}
                             0 & 0 \\
                             v_{L,R}e^{i\alpha_{L,R}} & 0 
                           \end{array} \right) \,, \,\ \        
\langle \phi_1 \rangle = \left( 
                            \begin{array} {ccc}
                             k_1 e^{i \alpha_1} & 0 \\
                             0 & k_2 e^{i \alpha_2} 
                             \end{array} \right)  \,.
\label{vevs}
\eeq
In order to have a phenomenologically viable model the 
VEVs must be hierarchical, 
\beq
|v_L|&\ll& k\ll|v_R|\,, \ \
k=\sqrt{|k_1|^2+|k_2|^2}\sim 250\GeV
\label{hierV}\,.
\eeq
If the ratio $r\equiv|k_2/k_1|$ is of order unity then in general
no splitting is obtained between the Dirac masses of the two
components of the SU(2)$_L$ doublets.
This causes phenomenological problems both in the lepton sector (see discussion below)
and in the quark sector \cite{Eck,Bar1,Ball1}.
Thus, we assume below that $r$ is a small number.   

Given the VEV hierarchy of \eq{hierV} the light neutrino
mass matrix, $M^\nu_l$, is given by:
\beq
M^\nu_l\approx (M_\nu^D)^T (M_R^{Maj})^{-1} M_\nu^D+M_L^{Maj}
\,,
\label{Mnulight}
\eeq
with $M_\nu^D$ being the Dirac neutrino mass and $M_{R}^{Maj}$ 
[$M_{L}^{Maj}$]
being the Majorana mass matrix for the right [left] handed neutrinos.
The RHS of \eq{Mnulight} contains two terms. The first, 
$M^\nu_{See}\equiv (M_\nu^D)^T (M_R^{Maj})^{-1} M_\nu^D$,
is related to the seesaw mechanism\cite{SeeSaw}. It 
depends on the scale of the Dirac neutrino masses and on $v_R$. The second,
$M_L^{Maj}$, depends on $v_L$.
In order to estimate the magnitude of $M^\nu_l$ 
we consider them separately.


\subsection{The natural value of $M^\nu_{See}$ and $M_{L}^{Maj}$}

\subsubsection{$M^\nu_{See}$}
In general, without further assumptions, the scale of
the charged lepton and the neutrino Dirac masses is roughly the
same. 
Therefore, neglecting mixing, the value of the tau neutrino mass 
or $(M^\nu_{See})_{33}$ is simply
given by the well known seesaw formula \cite{SeeSaw}:
\beq
(M^{\nu}_{See})_{33}\sim (m_\tau)^2/v_R= \Oc(0.1 \MeV)
\,.
\label{See}
\eeq
Such a high value is well above the bound given in 
\eq{nu}.


\subsubsection{$M^{maj}_L$}

The scalar potential of the MLRSM is given by
\footnote{There is a redundancy in different terms 
in the potential the way it is written in
eqs. (\ref{Vphi}-\ref{VDelphi}). 
It does not, however, affect
our analysis.},
\beq
V(\phi,\Delta)=V_{\phi}+V_{\Delta}+V_{\phi\Delta}\,,
\label{Vpot}
\eeq
where
\beq
{V}_\phi & = & -\mu^2_{ij}  \mbox{Tr}(\phi^\dagger_i \phi_j ) \, 
+ \,
 \lambda_{ijkl} \,  \mbox{Tr}( \phi^\dagger_i \phi_j )  
 \, \mbox{Tr}(\phi^\dagger_k  \phi_l)\,,
\label{Vphi}
\eeq

\beq
{V}_\Delta  & = & -\mu^2_i  \, \mbox{Tr}(\Delta_i \Delta_i^\dagger )
 \, + \,
 \rho_{ij} \, \mbox{Tr} (\Delta_i \Delta_i ) 
\mbox{Tr} (\Delta_j^\dagger \Delta_j^\dagger )
\label{VDel}
\,,
\eeq
with $\Delta_{1,2}=\Delta_{L,R}$ and
\beq
V_{\phi \Delta} & = &
\alpha_{ij} \, \mbox{Tr} (\phi^\dagger_i \phi_j ) 
\mbox{Tr} (\Delta_k \Delta_k^\dagger )  \,
+ \,   \, \mbox{Tr}(\beta_{ij}\Delta_L^\dagger \Delta_L\phi_i \phi^\dagger_j 
+\beta'_{ij}\Delta_R^\dagger \Delta_R\phi^\dagger_i \phi_j ) 
\nonumber\\ 
&+& \gamma_{ij}  \,
\mbox{Tr}(\Delta_L^\dagger \phi_i\Delta_R\phi^\dagger_j) 
\,.
\label{VDelphi}
\eeq 
It leads to the following relation \cite{LRSM,WB,Gun1,Gun2}: 
\beq
|v_R v_L|=a |k_1|^2(1+O(|r|))
\,,
\label{rel}
\eeq
where $a$ is a function of the various couplings.
Without fine tuning $a$ is expected to be of order unity.   
Substituting the values of the relevant 
VEVs given in \eq{vR} and \eq{hierV}
into \eq{rel} leads to a rather high value of $v_L$,
\beq
v_L = \Oc(1\GeV)
\label{v_L}
\,.
\eeq
Such a high value of $v_L$, with the triplet Yukawa couplings of order unity,  
yields light neutrino masses of the order of 1 GeV which
is above the direct experimental bound~\cite{PDG,Beta}.

To conclude, with $v_R\lsim15 \TeV$ and the Yukawas of the Higgs triplet
of $\Oc(1)$, we expect $m_{\nu_\tau}=\Oc(0.1-10^3\MeV)$, which 
is well above the eV scale.


\section{A MLRSM with a U(1) Symmetry Model}\label{toy}
Below we show how an additional $U(1)$ symmetry can solve
the problem of too massive active neutrinos discussed above.
For our demonstration we consider a model with only third
generation leptons.
Since the seesaw mechanism relates the heaviest charged lepton to
the heaviest neutrino, our model deals with the most severe
phenomenological problem.

The low energy effective theory with the additional U(1) symmetry
is assumed to be broken by a small parameter, $\eps$.
Thus, various terms in the low energy effective 
Lagrangian are  suppressed by powers of $\eps$,
\beq
{\cal L}^{eff}&=& {\cal L}_0+
{\eps}^{| Q(L_R)-Q(L_L)+Q(\phi_1) |} f \bar L_L \phi_1 L_R 
+ {\eps}^{| Q(L_R)-Q(L_L)-Q(\phi_1) |}g \bar L_L \phi_2 L_R
\nonumber \\
&+&{\eps}^{| Q(\Delta_L)+2Q(L_L) |}h(\Delta_L L_L L_L+\Delta_R L_R L_R)
+V(\phi,\Delta) +h.c.
\label{Lcorr}
\,,
\eeq
where we redefined the various Yukawa couplings so that $f,g,h=\Oc(1)$.
The structure of $V(\phi,\Delta)$ 
can be easily deduced from eqs.
(\ref{Vpot}-\ref{VDelphi}). To our consideration it is
important only to focus on the $\alpha_{ij},\beta_{ij}$ and
$\gamma_{ij}$ terms. We redefine them so that now $\alpha_{ij},\beta_{ij},\gamma_{ij}=\Oc(1)$,
\beq
{\alpha}_{ij},{\beta'}_{ji},{\beta}_{ij} \Rightarrow
\left\{
\begin{array}{ll}
(\alpha_{ij},\beta'_{ji},\beta_{ij}) \qquad & i=j \\
(\alpha_{ij},\beta'_{ji},\beta_{ij})
{\eps}^{2 | Q(\phi_1) |}   & i\neq j \\
\end{array}
\right.
\,
\label{sets1}
\eeq
and
\beq
{\gamma}_{ij}  \Rightarrow
\left\{
\begin{array}{ll}
 \gamma_{ij}{\eps}^{| Q(\Delta_R)-Q(\Delta_L) |}
 & i=j  \\
 \gamma_{ij} 
{\eps}^{| Q(\Delta_R)-Q(\Delta_L)+2Q(\phi_1) |}
&i=1 \quad j=2 \\
 \gamma_{ij}
{\eps}^{| Q(\Delta_R)-Q(\Delta_L)-2Q(\phi_1) |} 
 & i=2 \quad j=1 
\end{array}
\right.
\,.
\label{sets2}
\eeq
The most general VEVs that the fields can develop are given in
\eq{vevs}. Using,
however,
SU(2)$_L\,\times\,$SU(2)$_R$ transformations of the form
\beq
U_{L,R} =  \left(  \begin{array} {ccc}
                             e^{i \theta_{L,R}} & 0 \\
                             0 &e^{-i \theta_{L,R}}   
                             \end{array} \right)  
\eeq
one can bring $v_{L}$  and $v_R$ to be real.
Thus, the effective potential acquires the following form at the minimum:
\beq
V(\langle\phi_{1,2}\rangle,\langle\Delta_{R,L}\rangle)& =&
 - \mu^{2}_{3} (v_{L}^{2} + v_{R}^{2} ) + 
\frac{\rho}{4}(v_{L}^{4} + v_{R}^{4} ) + \frac{\rho'}{2} v_{L}^{2}  v_{R}^{2}
+ (v_{L}^{2} + v_{R}^{2} ) [ ( \alpha_{11} + \alpha_{22} + \bar\beta_{11})
| k_{1}|^{2} 
\nonumber\\ 
&+&
( \alpha_{11} + \alpha_{22} + \bar\beta_{22}) | k_{2}|^{2} 
+ {\eps}^{| 2Q(\phi_1) |} (4 \alpha_{12}+2 \beta_{12} )
\Re(k_{1} k_{2}^*) ] 
\nonumber\\ 
&+&
 2 v_{L} v_{R} 
[{\eps}^{| Q(\Delta_R)-Q(\Delta_L) |}( \gamma_{11} + \gamma_{22}) 
\Re(k_{1} k_2^*)  \nonumber\\
&+& {\eps}^{| Q(\Delta_R)-Q(\Delta_L)+2Q(\phi_1) |} \gamma_{12}
| k_{1} |^{2}  
+{\eps}^{| Q(\Delta_R)-Q(\Delta_L)-2Q(\phi_1) |}\gamma_{21}
| k_{2} |^{2}] 
\nonumber\\ 
&+& {\rm terms\ which\ depend\ only\ on}
\ \ k_1 \,, k_2 \,  \ ,   
\label{potential}
\eeq
where $\rho = 4 (\rho_{11} + \rho_{22}) \,, \rho' = \rho_{12}+\rho_{21} \,$
and $\bar\beta_{ij} =\beta_{ij}+\beta'_{ij}$.
As discussed above we assume $r,\eps \ll 1$. 
Therefore the potential $V(\phi,\Delta)$  
is approximately given by:
\beq
V(\langle\phi\rangle,\langle\Delta\rangle) &\approx& 
 - \mu^{2}_{3}(v_{R}^{2}+v_{L}^{2})  + 
\frac{\rho}{4}(v_{R}^{4}+v_{L}^{4})  + \frac{\rho'}{2}v_{R}^{2}v_{L}^{2}
\nonumber\\ 
&+& \frac{\alpha}{2}(v_{R}^{2}+v_{L}^{2})  
 |k_{1}|^{2}  + \beta v_{L} v_{R} |k_{1}|^{2} \,,
\eeq
with 
\beq
\alpha &\approx& 2 ( \alpha_{11} + \alpha_{22} + \bar\beta_{11})\,, 
 \nonumber\\
\beta &\approx& 2[r^2 {\eps}^{|Q(\Delta_R-\Delta_L)+2Q(\phi_1)|} \gamma_{12}
+ {\eps}^{|Q(\Delta_R)-Q(\Delta_L)-2Q(\phi_1)|} \gamma_{21}
\nonumber\\
&+&
r\cos(\alpha_1-\alpha_2) {\eps}^{|Q(\Delta_R)-Q(\Delta_L)|}
(\gamma_{11} + \gamma_{22})]
\,.
\label{effect}
\eeq
Following the standard analysis of the minimization of the potential
(see {\it e.g.} \cite{LRSM,WB,Gun1,Gun2,Bar5}) we obtain:
\beq
v_L v_R = \gamma |k_{1}|^2 \,,
\label{gamma}
\eeq
with $\gamma = \beta/(\rho - \rho' )$. 
Thus, in our model the suppression related to $\gamma$ 
is roughly given by:
\beq
\gamma \sim  \left\{max[r^2 {\eps}^{|Q(\Delta_L)-Q(\Delta_R)-2Q(\phi_1)|} ,
r{\eps}^{|Q(\Delta_L)-Q(\Delta_R)|} ,
{\eps}^{|Q(\Delta_L)-Q(\Delta_R)+2Q(\phi_1)|}]\right\}
\label{gammasup}
\,,
\eeq
which can naturally bring $v_L$ to below the eV scale
as required by \eq{nu}.

In our model the lepton masses are given by:
\beq
m_{\nu_{\tau}}^D &=& \frac{|k_1|}{\sqrt{2}}(f{\eps}^{|Q(L_L)-Q(L_R)-Q(\phi_1)|} +
rg{\eps}^{|Q(L_L)-Q(L_R)+Q(\phi_1)|}e^{i\theta})  \nonumber\\
m_{\tau}^D &=& \frac{|k_1|}{\sqrt{2}}(g{\eps}^{|Q(L_L)-Q(L_R)+Q(\phi_1)|}
+ rf{\eps}^{|Q(L_L)-Q(L_R)-Q(\phi_1)|}e^{-i\theta})
\label{masses}
\eeq
where we used a phase convention in which $v_R , k_1$ are real.

To make our discussion concrete we choose the following
set of charges:
\beq
Q(\Delta_L)=-Q(\Delta_R)=6\,,\ Q(\phi_1)=-Q(\phi_2)=2\,, \
 Q(L_L)=-Q(L_R)=3
\label{Qs}
\,.
\eeq
Assuming ${\eps}^4 \sim r \sim 1/250$, $v_R \sim
15 \TeV$, $k_1 \sim 250 \GeV$ we get
\beq 
v_L \sim {\eps}^{16} k_1^2/v_R\sim 0.1\eV
\label{v_La}
\,,
\eeq
\beq
m_\tau \sim \eps^4 k_1 g \sim 1 \GeV  \,,  \ 
m_{\nu_\tau}^D \sim \eps^4 k_1 |\eps^4 f+r g e^{i\theta}| \sim 1 \MeV 
\label{m^D}
\,,
\eeq
and
\beq
m_{\nu_{\tau}} \sim  h v_L - 
(m_{\nu_{\tau}}^D)^2/v_R \sim 0.1 \eV
\label{m_nutau}
\,.
\eeq

To summarize, we showed that, in principle, 
using an additional U(1) symmetry and 
under the assumption that $r$ is of $\Oc(10^{-2}-10^{-3})$,
one can construct a model in which $v_R$ is of
$\Oc(15\TeV)$, the neutrino masses are below $\Oc(1\eV)$ and
$m_\tau\sim 1\GeV$.
In that sense, models in which the typical scale of LRS breaking is
relatively low might still be natural.


\section{Can a U(1) Horizontal Symmetry Account for the recent SN Data?}\label{U1}

We examine below the possibility that the structure of the lepton
flavor sector is explained solely by the above U(1) symmetry, which
acts on the different flavors as an horizontal symmetry.
In particular we focus on the mixing between the first two
generations and investigate whether a large mixing angle (LMA) 
is obtained by the model, 
as strongly favored by the recent data from the 
SN experiments \cite{SNO,Analysis}.
We assume for simplicity that all the
right handed neutrinos are heavy. 
Thus our analysis does not apply to models of four light
neutrinos. We also neglect CP violation (CPV) in the lepton sector.

The charged lepton mass matrix, $M^{cl}$ is given by:
\beq
M^{cl}\sim{k\over\sqrt2}\pmatrix
{{\eps}^{|Q_3+{Q}^L_{13}+{Q}^R_{13}|}&{\eps}^{|Q_3+{Q}^L_{23}+{Q}^R_{13}|}&
{\eps}^{|Q_3+{Q}^R_{13}|}\cr
{\eps}^{|Q_3+{Q}^R_{23}+{Q}^L_{13}|}
&{\eps}^{|Q_3+{Q}^L_{23}+{Q}^R_{23}|} & 
{\eps}^{|Q_3+{Q}^R_{23}|}\cr
{\eps}^{|Q_3+{Q}^L_{13}}| &{\eps}^{|Q_3+{Q}^L_{23}|} & {\eps}^{|Q_3|}
}
\label{3gen}
\,,
\eeq
with $Q_i=Q(L_R^i)-Q(L_L^i)+Q(\phi_1)$, ${Q}^R_{ij}=Q(L_R^i)-Q(L_R^j)$
and  ${Q}^L_{ij}=Q(L_L^j)-Q(L_L^i)$.

Within the LRSM
the Dirac mass matrices are hermitian. Therefore,
\beq
|Q(L_R^i)-Q(L_L^j)+Q(\phi_1)|=|Q(L_R^j)-Q(L_L^i)+Q(\phi_1)|
\label{U1Qrel}
\,,
\eeq
with $i=1,2,3$. 
In the appendix we show that since in our framework the Dirac mass
matrices are both hermitian
and hierarchical then the following relation is obtained:
\beq
{Q}^R_{ij}={Q}^L_{ij}\equiv{Q}_{ij}
\label{hermit1}
\,.
\eeq
In addition, from \eq{m^D} and \eq{m_nutau} we learn that 
in order to obtain neutrino masses
below the eV scale, we must have:
\beq\label{upper}
m_{\nu_i}^D\lsim m_\tau/100
\,,
\eeq
with $m_{\nu_i}^D$, $i=1,2,3$ are the eigenvalues of the
neutrino Dirac mass matrix.

Within our model, the suppression of the neutrino Dirac 
masses is achieved by aligning the sign of the U(1) charges of the
lepton doublets with the one of $\phi_1$.
This yields a suppression  of $\eps^{2|Q(\phi_1)|}$
of the neutrinos Dirac masses compared with the charged lepton ones.
This is demonstrated in 
eqs. (\ref{masses},\ref{Qs},\ref{m^D}), that uses, 
$sign[Q(L_R^3)-Q(L_L^3)]=sign[Q(\phi_1)]$.
Though a charge assignment in which $sign(Q_3)\neq sign(Q_{1,2})$
can, in principle, lead to a viable (hermitian and hierarchical)
charged lepton mass matrix, it yields an enhancement of
$\eps^{-2|Q(\phi_1)|}= \Oc(100)$ to the corresponding entries in
the neutrino mass matrix. It is enough that the first generation is assigned
an opposite charge, $sign(Q_3)=sign(Q_2)\neq sign(Q_1)$, 
to typically have $m_{\nu_1}^D$ of the
order of 100 \MeV. This  leads to a violation of the constraint in \eq{upper}.
The same clearly holds for the other generations.
Thus we must have:
\beq
sign(Q_1)=sign(Q_2)=sign(Q_3)
\label{signs}
\,.
\eeq
The hierarchy in the charged lepton masses
and eqs. (\ref{U1Qrel},\ref{hermit1},\ref{signs}) 
lead to the following relations \cite{GN}:
\beq
\left|{(M^{D}_{\nu})_{12}\over(M^{D}_{\nu})_{22}}\right|&\sim&\left|{M^{cl}_{12}\over M^{cl}_{22}}\right|\sim \sqrt{m_e\over
m_\mu}\ , \ \
\left|{(M^{D}_{\nu})_{23}\over(M^{D}_{\nu})_{33}}\right|\sim\
\left|{M^{cl}_{23}\over M^{cl}_{33}}\right|\sim \sqrt{m_\mu\over
m_\tau}\ , \nonumber\\
\left|{(M^{D}_{\nu})_{13}\over(M^{D}_{\nu})_{33}}\right|&\sim&
\left|{M^{cl}_{13}\over M^{cl}_{33}}\right|\sim \sqrt{m_e\over
m_\tau}
\label{hier}
\,.
\eeq
From \eq{hier} we learn that the 
the charged lepton mass
matrix typically gives very small mixing angles
in the (12) and (13) planes.
Thus LMA for the SN can only come from the 
light neutrino mass matrix. 

With horizontal U(1) symmetries there are two typical structures 
of neutrino mass matrix that may account for both LMA in the
(12) and (23) planes \cite{Barb,GNS}.
One of them involves a special structure of the $3\times3$ mass
matrix. This structure cannot be reduced to a $2\times2$ block 
matrix description. 
In the appendix we show that in our case
this type cannot lead to LMA in the (12) plane.
In the second structure, the LMA of the SN 
comes from the entries of the 2$\times$2 block mass matrix
related to the first two generations.
Below we separately investigate $M^\nu_{See}$ and $M_L^{Maj}$
and show that LMA in the relevant 2$\times$2 block of the mass matrix
cannot be obtained in our framework.

\subsection{Mixing in $M^\nu_{See}$}

Consider the relevant 2$\times$2 block in $M^\nu_{See}$. Requiring
a LMA between the first two generations 
is translated to the following condition:
\beq 
(M^\nu_{See})_{12}\gsim max\left[(M^\nu_{See})_{11},(M^\nu_{See})_{22}\right]
\label{bigmix}
\,.
\eeq
Since the neutrino Dirac mass matrix, $M_\nu^D$, is quasi diagonal
it is clear that, to satisfy the condition of \eq{bigmix},
the following ratio between the entries of
$(M_R^{Maj})^{-1}$ should hold:
\beq 
R^R_{12}\equiv\left|{(M_R^{Maj})^{-1}_{22}\over(M_R^{Maj})^{-1}_{12}}\right| 
\lsim {m_e\over m_\mu}
\label{mixinv}
\,.
\eeq
In the appendix we show that the ratio $R^R_{12}$ is larger than
$\sqrt{m_e\over m_\mu}$ and the condition of \eq{mixinv} cannot be satisfied.
Therefore LMA in the relevant block of $M^\nu_{See}$ is impossible.
The above conclusion can be explained as follows.
Within our framework strong hierarchy in the neutrino Dirac matrices,
required by the hierarchy of the charged lepton masses, is inevitable.
Such strong hierarchy suppresses any off-diagonal dominance
that can be obtained in $(M_R^{Maj})^{-1}$. Therefore 
order one mixing in $M^\nu_{See}$ cannot be obtained.

\subsection{Mixing in $M_L^{Maj}$}\label{MixM_L}

$M_L^{Maj}$ is given by:
\beq
M_L^{Maj}=v_L
\pmatrix
{
{\eps}^{2|Q'_3+{Q}_{13}|}&{\eps}^{|2Q'_3+{Q}_{23}+{Q}_{13}|}&
{\eps}^{|2Q'_3+{Q}_{13}|}\cr
{\eps}^{|2Q'_3+{Q}_{23}+{Q}_{13}|}&
{\eps}^{2|Q'_3+{Q}_{23}|}&{\eps}^{|2Q'_3+{Q}_{23}|}\cr
{\eps}^{|2Q'_3+{Q}_{13}|}&{\eps}^{|2Q'_3+{Q}_{23}|}&{\eps}^{2|Q'_3|}
}
\label{mL}
\,,
\eeq
where $2Q'_3=2Q(L_R^3)+Q(\Delta_R)$ and we used the fact that within
LRSM $|M^{Maj}_L/v_L|=|M^{Maj}_R/v_R|$.

It is evident from \eq{mL} that a proper charge assignment for
$Q(\Delta_{R,L})$, 
\beq
Q(\Delta_R)\simeq-2Q(L_R^3)-{Q}_{23}-{Q}_{13}\,; \ 
Q(\Delta_L)\simeq-2Q(L_L^3)+{Q}_{23}+{Q}_{13}
\label{QDelR}
\,,
\eeq
yields dominance of $(M_L^{Maj})_{12}$ over the diagonal entries
$(M_L^{Maj})_{11}$ and $(M_L^{Maj})_{22}$.
However, the condition for large mixing, 
$(M^\nu_L)_{12}\gsim max\,[(M^\nu_L)_{11},(M^\nu_L)_{22}]$, together
with \eq{Mnulight} imply that one needs also to compare 
$(M^\nu_L)_{12}$ to $(M^\nu_{See})_{11}$ and $(M^\nu_{See})_{22}$.
As shown in \eq{gamma}, the scales of $M^{Maj}_L$
and $M^\nu_{See}$  are not independent in the LRSM. In particular,
the higher the charges chosen for the triplet fields the smaller the product 
$v_L v_R$ becomes. Thus for a fixed value of $v_R$ the contribution of 
$M^{Maj}_L$ to $M^\nu_l$ might become negligible.

In our framework we saw that $(M^\nu_{See})_{22}$ is
expected to be dominant compared with the other 
entries of $(M^\nu_{See})_{ij}$, $i,j= 1,2$. 
Therefore we compare it to
$(M_L^{Maj})_{12}$ assuming that the condition of \eq{QDelR} holds.
Given \eq{QDelR} the determinant of $M_R^{Maj}$ is given by:
\beq  
det(M_R^{Maj})\sim {v_R^3 {\eps}^{|{Q}_{13}+{Q}_{23}|}}
\label{detR}\,.
\eeq
Using eqs. (\ref{Mnulight},\ref{invmR},\ref{detR}) we find that:
\beq
(M^\nu_{See})_{22}\sim {(m^D_{\nu_\mu})^2\over v_R {\eps}^{|{Q}_{12}|}}
\,.
\label{MS22}
\eeq
Hence, the ratio $(M_L^{Maj})_{12}/(M^\nu_{See})_{22}$
is given by:
\beq
R^L_{12}&\equiv&(M_L^{Maj})_{12}/(M^\nu_{See})_{22}\approx
{v_R v_L \over (m^D_{\nu_\tau})^2} 
{\eps}^{|{Q}_{12}|-4|{Q}_{23}|}
\label{ratio}
\,.
\eeq
Thus, LMA is induced by $M_L^{Maj}$ if,
\beq
R^L_{12}\gsim1
\label{RL}
\,.
\eeq 
In the appendix we show that, for the ratio $R^L_{12}$ to be larger
than unity, it is
required that $v_R\gsim10^9$ which is
irrelevant for our discussion. Therefore obtaining LMA
in the (12) block is impossible. 

In other words, we saw that in order to produce LMA in $M_L^{Maj}$ one is driven
to assign relatively high charges for $\Delta_{L,R}$ as shown in \eq{QDelR}.
The high charges of $\Delta_{L,R}$ induce suppression of the product
$v_L v_R$, as seen in \eq{gamma} and \eq{gammasup}. This suppression
in its turn yields suppression of the ratio $R^L_{12}$.
 
\subsection{Conclusion}
From the  above analysis we learn that a MLRSM with the additional horizontal
U(1) symmetry discussed above cannot account for the LMA solution of 
the SN problem in a natural way.
We focus on this point since
the combind data from the SNO and Super-Kamiokande 
experiments~\cite{SNO,Nuexp}
disfavors the SMA solution of the SN problem~\cite{Analysis}. 
Though we do not show it explicitly in the text, a proper charge
assignment for the various fields, {\it e.g.}\,, 
$Q(\Delta_L)=-Q(\Delta_R)=8\,,\ Q(\phi_1)=2\,, \
Q(L_L^{1,2,3})=-Q(L_R^{1,2,3})=3,4,6$, and an appropriate choice of the
ratio $\eps^4/r$ can produce a viable model of the
lepton sector without any fine tuning. 
Such a model yields order one mixing for the atmospheric neutrinos
and the SMA solution to the SN problem.

\section{Discussion and Conclusion}\label{conclusion}

A generic LRSM with the left-right symmetry (LRS) 
breaking scale $v_R\lsim15\TeV$ and natural value for Yukawa
Higgs couplings leads 
to a severe problem in the lepton sector: 
The light neutrino masses are in fact rather heavy.
The problem arises from two different, generically, 
independent parts of the model:\\  
1. $v_L$, the triplet VEV is relatively high, $v_L\sim k^2/v_R= \Oc(1\GeV)$, 
which leads to a large eigen values
of the Majorana mass matrix for the left-handed neutrinos. \\
2. The scale of the neutrino mass matrix obtained by the seesaw 
mechanism is large, with the mass of the heaviest neutrino 
of the order $m_\tau^2/v_R= 0.1\MeV$, provided that the 
tau neutrino Yukawa
couplings are similar to those of the tau.\\

In this context we demonstrated that, within a minimal LRSM, 
the two problems are solved using an additional U(1)
symmetry, when the ratio between the two VEVs of the
bi-doublet neutral fields is assumed to be small, of the order 
$10^{-2}-10^{-3}$.

We further showed that, if we assume that the same U(1) symmetry
explains the charged lepton flavor hierarchy, 
small mixing angle is obtained between the first two generations. 
Nevertheless, there is
no reason that a viable model, which combines the above
U(1) symmetry with additional symmetries, cannot be constructed. 

This work is motivated by two reasons: 
First, the idea that the fundamental scale of gravity 
might be much smaller than
the Planck scale was recently proposed and is, at present, 
consistent with all of our experimental knowledge~\cite{LED}.
Our work demonstrates
how a LRSM, in which the fundamental scale of gravity 
(and other NP sources) is relatively low, can
account in a natural way for 
the smallness of neutrino masses. 
To the best of our knowledge no such complete LRSM was 
considered so far in the literature~\cite{Ma}. 

Our second motivation is related to the fact that LRSM with
spontaneous CP violation 
(SCPV)~\cite{Eck,WB} were recently analyzed
in the literature~\cite{Bar1,Ball1,Bar5,Bar2,Bar3,Bar4,Ball2,BP2} and 
found to have a considerable agreement with all the CP conserving
experimental data. 
It was also shown there that constraints from the quark sector
typically require $r$ to be of 
$\Oc(10^{-2}-10^{-3})$\cite{Ball1,Bar3,Ball2,BP2},
and that $v_R$ should be of $\Oc(10\TeV)$\cite{Ball1,Ball2}.

In that context it is important to note that MLRSM with SCPV 
cannot account for the observed CPV in the quark
sector~\cite{WB,Gun2,Bar5}. Recently it was also shown in \cite{Bar5}
that the CP properties
of the vacuum of the MLRSM with SCPV are strongly connected to the
physical Higgs fields mass spectrum. This means 
that the above problem of the MLRSM with SCPV 
might be shared by many LRSM with SCPV. 
To what extent this problem is really general  
is yet to be investigated. Therefore LRSM with SCPV
should not be discarded.

Finally we remark that the recent result of BaBar\cite{BaBar} and Belle
\cite{Belle} experiments exclude any LRSM with SCPV in which
the quarks fields couples only to a single
bidoublet field~\cite{Ball1,Ball2,BP2,Beneke}.

\acknowledgements
 
We thank Yossi Nir for helpful discussions and comments
on the manuscript.

\newpage

\appendix

\section{}

\subsection{The Values of ${Q}^{LR}_{ij}$}

In this subsection we prove \eq{hermit1}, that is, 
${Q}^{R}_{ij}={Q}^{L}_{ij}$.
Consider a two generation picture and suppose that
${Q}^{R}_{12}\neq{Q}^{L}_{12}$ and that, without 
loss of generality, $sign({Q}^{R}_{12})=sign(Q_2)$, with $Q_2=Q(L_R^2)-Q(L_L^2)+Q(\phi_1)>0$.
The only way to ensure that the product $M_{12}\cdot M_{21}$
is of the order of $m_1 m_2$, as required by
the hierarchy of the charged lepton masses, is when ${Q}^{L}_{12}$ is given by:
\beq
{Q}^L_{12}\simeq -({Q}^R_{12}+2Q_2)
\label{kind2app}
\,.
\eeq
In this case it is easy to see that $M_{11}$ is of
the same order as $M_{22}$ which does not allow for $m_1\ll m_2$.
Therefore we conclude that $\eq{hermit1}$ must be satisfied.

\subsection{Mixing in a Non-reducible 3$\times$3 Majorana Neutrino Mass Matrix}
In ref \cite{Barb} it was shown how an $L_e-L_\mu-L_\tau$
symmetry can lead to LMA both for the AN and for the SN.
However it was assumed in \cite{Barb} that the hierarchy in the
charged lepton masses comes from additional flavor symmetry. Furthermore
the 2$\times$2 block related to the second and third generations in the
Dirac neutrino mass matrix was assumed to be non-hierarchical. 
In our case we do not allow for additional continuous symmetries, therefore
an $L_e-L_\mu-L_\tau$ symmetry cannot be realized, and the Dirac
matrices obey the structure given in \eq{hier}.
Nevertheless we want to consider the case that neutrino Majorana mass
matrices have an approximate $L_e-L_\mu-L_\tau$
structure\footnote{This can be achieved, {\it e.g.}\,, using the following charge
assignment, $Q(\Delta_L)=-Q(\Delta_R)=8.5\,,\ Q(\phi_1)=2\,, \
Q(L_L^{1,2,3})=-Q(L_R^{1,2,3})=3,4,11/2$, this leads to electron parity
symmetry which we assume to be broken by another small parameter
$\eps'\ll \eps$.}:
\beq
M^{Maj}_{L,R}\approx v_{L,R}\pmatrix{\eps'&a&1\cr a&\eps'&\eps'\cr 1&\eps'&\eps'}
\,,\label{MRapp}
\eeq
with $\eps'\ll|a|\lsim1$. 

Applying the analysis of~\cite{Barb} to our model 
and assuming for simplicity a diagonal Dirac mass matrix, typically yields
the following structure for $M^\nu_{See}$:
\beq
M^\nu_{See}\sim {\eps^{4|Q(\phi_1)|}\over v_R}\pmatrix{\eps' m_e^2 & m_\mu m_e &m_e m_\tau\cr
m_\mu m_e & \eps' m_\mu^2&0\cr
m_\tau m_e &0&{m_\tau^2\over\eps'} }
\,,
\label{33}
\eeq
where we assumed also that $r\sim{\eps}^{2|Q(\phi_1)|}$.
As clearly seen from \eq{33} to have LMA for the SN one needs
$\eps'\lsim {m_e\over m_\mu}$. This means that
the heavy neutrino mass is enhanced by the inverse of the same factor
which clearly violates \eq{upper}. In other worlds, LMA cannot be
induced by $M^\nu_{See}$.

In the case where $\eps'$ is relatively large, 
as required by \eq{upper}, LMA also
cannot be induced by $M^{Maj}_L$.
To see that one needs to verify that
${(M^{Maj}_L)_{12}\over (M^\nu_{See})_{22}}=R^L_{12}\ll1$.
This ratio is investigated in subsection \ref{MixM_L} and in
subsection \ref{aMixM_L}, below, in a similar situation. 
Applying the same analysis to the present case one finds that
the value of the ratio $R^L_{12}$ 
is enhanced by a factor of $1/\eps'$ relative to
the corresponding values given in eqs. (\ref{Q1},\ref{Q2},\ref{Q3}). 
The mild enhancement, of $\Oc(1/\eps')$,
still yields $R^L_{12}\ll1$ from which we learn 
that the mixing angle induced by $M^{Maj}_L$ are small.  

Thus the above scenario cannot lead to LMA between the first two generations.

\subsection{Mixing in $M^\nu_{See}$}

Below we show that the ratio $R^R_{12}$ defined in \eq{mixinv} is always larger than
$\sqrt{m_e\over m_\mu}$ and therefore \eq{mixinv} cannot be satisfied.
In order to do so we focus on the structure of $(M_R^{Maj})^{-1}$,
\beq
(M_R^{Maj})^{-1}={1\over v_R}
\pmatrix
{
{\eps}^{2|Q'_3+{Q}_{13}|}&{\eps}^{|2Q'_3+{Q}_{23}+{Q}_{13}|}&
{\eps}^{|2Q'_3+{Q}_{13}|}\cr
{\eps}^{|2Q'_3+{Q}_{23}+{Q}_{13}|}&
{\eps}^{2|Q'_3+{Q}_{23}|}&{\eps}^{|2Q'_3+{Q}_{23}|}\cr
{\eps}^{|2Q'_3+{Q}_{13}|}&{\eps}^{|2Q'_3+{Q}_{23}|}&{\eps}^{2|Q'_3|}
}^{-1}
\label{invmR}
\,.
\eeq
From \eq{invmR} we find that the  
ratio $R^R_{12}$ is given by:
\beq 
R^R_{12}=\left|{
{\eps}^{2|Q'_3+{Q}_{13}|+2|Q'_3|}-{\eps}^{2|2Q'_3+{Q}_{13}|}
\over
{\eps}^{|2Q'_3+{Q}_{13}+{Q}_{23}|+2|Q'_3|}-{\eps}^{|2Q'_3+{Q}_{13}|
+|2Q'_3+{Q}_{23}|}
}\right| 
\label{22/12}
\,.
\eeq
To see whether \eq{mixinv} can be satisfied 
we are interested in the minimal value that the ratio $R^R_{12}$ can have.
To find it we divide it into separate ranges of $Q'$ and calculate
the minimal value of $R^R_{12}$ in each region.
Without loss of generality we assume that $Q'$ is non-negative,
the result we get at the end of the discussion is valid for any value
of $Q'$.
\begin{itemize}
\item ${Q}_{13}>-Q'$:\\
In that case all the terms in the exponents of \eq{22/12} are positive and it
is easy to see that $R^R_{12}$ is given by:
\beq
R^R_{12}\sim {\eps}^{|{Q}_{12}|}\sim \sqrt{m_e\over m_\mu}
\,,
\label{it1}
\eeq
thus the condition of \eq{mixinv} is not satisfied.
\item $-2Q'\geq{Q}_{13}\geq-Q'$ and $-2Q'\geq{Q}_{13}+{Q}_{23}$ :\\
In that case only the exponent of the first term in the dominator is
negative and the ratio is given by:
\beq
R^R_{12}\sim max\,{\left[
{\eps}^{|2{Q}_{13}|}\,,\ {\eps}^{4Q'_3+2{Q}_{13}}\right]
\over
{\eps}^{4Q'_3+{Q}_{13}+{Q}_{23}}}
\geq  {\eps}^{|{Q}_{12}|} \sim \sqrt{m_e\over m_\mu}
\,,
\label{it2}
\eeq
where in the last line we used the relation ${Q}_{13}={Q}_{12}+{Q}_{23}$.
Therefore, the condition of \eq{mixinv} is  not satisfied.

\item  $-2Q'\geq{Q}_{13}\geq-Q'$ and $-2Q'>{Q}_{13}+{Q}_{23}$ :\\
In that case the exponent of the first term in the dominator and also
that of the first term in the denominator are 
negative and the ratio is given by:
\beq
R^R_{12}\sim {max\,\left[
{\eps}^{|2{Q}_{13}|}\,,\ {\eps}^{4Q'_3+2{Q}_{13}}\right]
\over max\,\left[{\eps}^{2|{Q}_{13}+{Q}_{12}|}\,,\ 
{\eps}^{4Q'_3+{Q}_{13}+{Q}_{23}}\right]}
\geq  {\eps}^{|{Q}_{12}|} \sim \sqrt{m_e\over m_\mu}
\,,
\label{it3}
\eeq
and the condition of \eq{mixinv} is  not satisfied.
\item  $-2Q'>{Q}_{13}$ and $-2Q'\leq{Q}_{23}$ :\\
In that case the exponent of the first term in the dominator and also
that of the first term in the denominator are 
negative. Therefore $R^R_{12}$ is given by:
\beq
R^R_{12}\sim {
{\eps}^{|2{Q}_{13}|}\over {\eps}^{2|{Q}_{13}+{Q}_{12}|}}
\geq  {\eps}^{|{Q}_{12}|} \sim \sqrt{m_e\over m_\mu}
\,,
\label{it4}
\eeq
and the condition of \eq{mixinv} is  not satisfied.
\item  $-2Q'>{Q}_{13}$ and $-2Q'>{Q}_{23}$ :\\
In that case the exponent of the first term in the dominator and also
that of the first term in the denominator are 
negative. Therefore the ratio is given by:
\beq
R^R_{12}\sim {max\,\left[
{\eps}^{|2{Q}_{13}|}\,,\ {\eps}^{-4Q'_3-2{Q}_{13}}\right]
\over max\,\left[{\eps}^{2|{Q}_{13}+{Q}_{12}|}\,,\ 
{\eps}^{-4Q'_3-{Q}_{13}-{Q}_{23}}\right]}
\geq  {\eps}^{|{Q}_{12}|} \sim \sqrt{m_e\over m_\mu}
\,,
\label{it5}
\eeq
and the condition of \eq{mixinv} is  not satisfied.
\end{itemize}
Thus we conclude that mixing of order one in $M^\nu_{See}$
between the first two generation cannot be achieved.


\subsection{Mixing in $M_L^{Maj}$}\label{aMixM_L}
Below we show that the ratio $R^L_{12}$, defined in \eq{ratio}, 
is larger than unity only if $v_R\gsim10^9$ which is
irrelevant for our discussion and therefore \eq{RL} cannot be satisfied.
The ratio $R^L_{12}$ is given by:
\beq
&R^L_{12}&\approx
{v_R v_L \over (m^D_{\nu_\tau})^2} 
{\eps}^{|{Q}_{12}|-4|{Q}_{23}|}
\sim {k_1^2\over (m^D_{\nu_\tau})^2} 
{\eps}^{|{Q}_{12}|-4|{Q}_{23}|}\times\nonumber\\
&&
max[r^2 {\eps}^{2|{Q}_{23}+{Q}_{13}+Q_3-Q(\phi_1)|} ,
r{\eps}^{2|{Q}_{23}+{Q}_{13}+Q_3|} ,
{\eps}^{2|{Q}_{23}+{Q}_{13}+Q_3+Q(\phi_1)|}]
\nonumber\\
&&\sim {\eps}^{|{Q}_{12}|-4|{Q}_{23}|}
{max[r^2 {\eps}^{2|{Q}_{23}+{Q}_{13}+Q_3-Q(\phi_1)|} ,
r{\eps}^{2|{Q}_{23}+{Q}_{13}+Q_3|} ,
{\eps}^{2|{Q}_{23}+{Q}_{13}+Q_3+Q(\phi_1)|}]
\over
max[r^2 {\eps}^{2|Q_3+Q(\phi_1)|} ,
r{\eps}^{|Q_3-Q(\phi_1)|+|Q_3+Q(\phi_1)|} ,
{\eps}^{2|Q_3-Q(\phi_1)|}]
}
\label{RLratio}
\,,
\eeq
where in the second line we used eqs. 
(\ref{gamma},\ref{gammasup},\ref{QDelR}) to simplify the above expression. 
To see whether the ratio $R^L_{12}$ could be equal or larger than unity 
we investigate \eq{RLratio} in four regimes relates to $Q(\phi_1)$. 
We assume that $Q_3$ and also ${Q}_{ij}$ are 
non-negative (since they carry the same sign according to our above
assumption). This assumption does not affect our final conclusion since
we do not impose any further assumptions on $Q(\phi_1)$:
\begin{itemize}
\item[(i)] $Q_3\geq Q(\phi_1)\geq0$ :\\
The ratio of \eq{RLratio} is given by,
\beq
R^L_{12}&\sim {\eps}^{3|{Q}_{12}|}
\cdot max[r^2, {\eps}^{2Q(\phi_1)},{\eps}^{4Q(\phi_1)}]\ll1
\,,
\label{Q1} 
\eeq
which means that in this case no large mixing is possible.

\item[(ii)] ${Q}_{23}+{Q}_{13}+Q_3 \geq Q(\phi_1)>Q_3$ :\\
The ratio of \eq{RLratio} is given by,
\beq
R^L_{12}&\sim {\eps}^{3|{Q}_{12}|}
{\eps}^{4Q_3}\ll1
\,,
\label{Q2} 
\eeq
which means that in this case no large mixing is possible.
\item[(iii)] $Q(\phi_1)>{Q}_{23}+{Q}_{13}+Q_3$ :\\
The ratio of \eq{RLratio} is given by:
\beq
R^L_{12}&\lsim {\eps}^{3|{Q}_{12}|} 
{\eps}^{|{Q}_{23}+{Q}_{13}|}\ll1
\,,
\label{Q3} 
\eeq
where we used the fact that we require $r \lsim {\eps}^{2|Q(\phi_1)|}$
as discussed in section \ref{toy}.
\Eq{Q3} implies that no large mixing is possible.

\item[(iv)] In items (i)-(iii) we considered cases with $Q(\phi_1)>0$.
Let us consider the case with negative charges for example consider
the range $-Q_3<Q(\phi_1)<0$ :\\
The ratio of \eq{RLratio} is given by,
\beq
R^L_{12}&\sim {\eps}^{3|{Q}_{12}|} r^{-2}
\,,
\label{Q4} 
\eeq
From \eq{Q4} we learn that maximal mixing is possible if
\beq 
r\lsim
{\eps}^{1.5|{Q}_{12}|}\sim 1/50
\,,
\label{large}
\eeq
which means that  large mixing is in principle possible.

However, as discussed above,
opposite charges of $\phi$  and $Q_3$ yield inverse hierarchy between the
neutrino and charge lepton Dirac masses, 
\beq 
{m^D_{\nu_i}\over m^D_i}\sim
{\eps}^{-2|Q\phi|}\gsim 1
\,.
\label{invhier1} 
\eeq
It means that in order not to produce too large neutrino masses 
$v_R$ is bounded from 
below:
\beq
v_R\geq {(m^{\nu_\tau}_D)^2\over 1\eV} \gsim   
{m_\tau^2\over 1\eV}\sim10^{9}\GeV   
\,.
\label{v_R1}
\eeq
Thus in this case large mixing  cannot be obtained
with a low value of $v_R$. 
A similar conclusion is clearly obtained for any negative value of 
$Q(\phi_1)$.
\end{itemize}




\begin{thebibliography}{99}
\bibitem{LRSM}
J.C. Pati and A. Salam, \prd{10} (1974) 275;
R.N. Mohapatra and J.C. Pati, \prd{11} (1975) 566 and 2558;
G. Senjanovi\'c and R.N. Mohapatra, \prd{12} (1975) 1502;
G.~Senjanovi\'c, Nucl. Phys. {\bf B153}  (1979) 334;
R.N. Mohapatra and G. Senjanovi\'c, \prl{44} (1980) 912; 
\prd{23} (1981) 165;
C. S. Lim and T. Inami,
Prog. Theor. Phys. {\bf 67} (1982) 1569.


\bibitem{GUT}
T. G. Rizzo and G. Senjanovi\'c, Phys. \ Rev. \ Lett. {\bf 46} (1981) 1315;
Phys. \ Rev.\ {\bf D24} (1981) 704; \prd{25} (1982) 235;
M. Fukugita, T. Yanagida and M. Yoshimura, 
Phys. \ Lett. {\bf B106} (1981) 183; 
R. N. Mohapatra, Fortsch.\  Phys. {\bf 31} (1983) 185. 

\bibitem{Cos}S.S. Gershtein and Ya. B. Zeldovich, 
JETP \  Lett. {\bf4} (1966) 120;
R. Cowsik and J. McClelland, Phys.\ Rev.\ Lett. {\bf29} (1972) 669; 
J.R. Primack and M.A.K. Gross, \aspha{0007165},
to appear in  
{\it Current Aspects of Neutrino Physics}, ed. D. O. Caldwell 
(Springer, Berlin Heidelberg 2000). 


\bibitem{PDG}
D.E. Groom {\it et. al.}, \epjc{15} (2000) 1. 

\bibitem{Beta}L. Baudis {\it et. al.}, Heidelberg-Moscow Double-Beta
Decay Exp.,
Phys.\ Rev.\ Lett. {\bf83} (1999) 41 \exb{9902014};
H.V. Klapdor-Kleingrothaus {\it et al.}, 
Heidelberg-Moscow Double-Beta Decay Exp.,
Talk presented at the third International Conference- DARK2000, 
to be published. in Proc. of DARK2000, Springer (2000), \pha{0103062}.



\bibitem{SNO}
Q.R. Ahmad {\it et al.}, SNO Collaboration, 
\prl{87}, 071301 (2001) \nucexb{0106015}.

\bibitem{Nuexp}B.T. Cleveland {\it et al.}, Astrophys. J. 
{\bf 496} (1998) 505;
W. Hampel {\it et al.}, GALLEX Collaboration, Phys.\ Lett. {\bf B447} (1999) 127;
J.N. Abdurashitov {\it et al.}, SAGE Collaboration, 
Phys. \ Rev. {\bf C 60} (1999) 055801 \asphb{9907113}; 
M. Altmann {\it et al.}, GNO Collaboration, Phys. \ Lett.
{\bf B490} (2000) 16 \exb{0006034};
S. Fukuda {\it et al.}, 
Super-Kamiokande Collaboration, Phys.\ Rev.\ Lett. {\bf 81} (1988) 4279; 
Phys.\ Rev.\ Lett. {\bf 86} (2001) 5651;
M. Apollonio {\it et al.}, CHOOZ Collaboration, Phys.\ Lett. 
{\bf B466} (1999) 415 \exb{9907037}.




\bibitem{Analysis}
A. Bandyopadhyay, S. Choubey, S. Goswami and Kamales Kar \pha{0106264}; 
J. N. Bahcall, M.C. Gonzalez-Garcia and C. Pena-Garay, \pha{0106258};
V. Barger, D. Marfatia and K. Whisnant \pha{0106207};
G.L. Fogli, E. Lisi, D. Montanino and A. Palazzo, \pha{0106247}. 


\bibitem{Eck}
G. Ecker, W. Grimus and H. Neufeld, \npb{247} (1984) 70;
G. Ecker and W. Grimus,  \npb{258} (1985) 328; 
Z. Phys. {\bf C30} (1986) 293; 
D. Chang, \npb{214} (1983) 435;
J. M. Frere {\it et. al.}, \prd{46} (1992) 337;
J. Basecq and D. Wyler,
Phys.\ Rev.\  {\bf D39} (1989) 870;
G.C. Branco and L. Lavoura, Phys.\ Lett.\  {\bf B165} (1985) 327.





\bibitem{Bar1} 
G. Barenboim, J. Bernabeu and M. Raidal,
Nucl. Phys. {\bf B478} (1996) 527 {[{\tt hep-ph/9608450}]}; 
Z.\ Phys.\ C {\bf 73} (1997) 321 \phb{9603379}.


\bibitem{Ball1}
P. Ball, J.M. Frere and J. Matias, \npb{572} (2000) 3 
{[{\tt hep-ph/9910211}]}.




\bibitem{SeeSaw}
M. Gell-Mann, P. Ramond and R. Slansky, in {\it Supergravity,} ed.
P. van Niewenhuizen and D. Freedman (North-Holland 1979);
T. Yanagida, in {\it Proceedings of the Workshop on the Unified
Theory and the Baryon Number in the Universe,} ed. O. Sawada and
A. Sugamoto (Tsukuba 1979);
R. N. Mohapatra and G. Senjanovi{\'c},
Phys. Rev. Lett. {\bf 44}, 912 (1980).


\bibitem{WB}J. Basecq, J. Liu, J. Milutinovic and L. Wolfenstein,
Nucl.\ Phys. {\bf B272} (1986) 145. 


\bibitem{Gun1}J.F. Gunion, J. Grifols, A. Mendez, B. Kayser and
F. Olness, 
\prd{40} (1989) 1546. 


\bibitem{Gun2}N.G. Deshpande, J.F. Gunion, B. Kayser and F. Olness, \prd{44}, 837 (1991). 
\bibitem{Bar5}G. Barenboim, M. Gorbahn, U. Nierste and M.
Raidal, \pha{0107121}. 


\bibitem{GN} Y Grossman and Y. Nir, Nucl. Phys. {\bf B448} (1995)
30 \phb{9502418}. 

\bibitem{Barb} R. Barbieri {\it et. al.}, JHEP {\bf 12} (1998) 98017 \phb{9807235}.

\bibitem{GNS}Y. Grossman, Y Nir and Y. Shadmi, JHEP {\bf 9810}
(1998) 007 \phb{9808355}. 


\bibitem{LED}
N. Arkani-Hamed, S. Dimopoulos and G. Dvali, Phys. Lett. {\bf
B429} (1998) 263 \phb{9803315}; 
I. Antoniadis, N. Arkani-Hamed, S. Dimopoulos and 
G. Dvali, Phys. Lett. {\bf B436} (1998) 257 \phb{9804398};
N. Arkani-Hamed and S. Dimopoulos \pha{9811353};
Z. Berezhiani and  Gia Dvali, Phys. Lett. {\bf B450} (1999) 24 \phb{9811378}; 
N. Arkani-Hamed, L. Hall, D. Smith and N. Weiner,
Phys. Rev. {\bf D61} (2000) 116003  \phb{9909326}.


\bibitem{Ma}Papers that discuss small neutrino masses induced by a
low-energy seesaw like mechanism in a non-LRS framework: 
E. Ma, Phys. Rev. Lett. {\bf86} (2001) 2502
\phb{0011121}; \pha{0107177}; \pha{0103278};
E. Ma, M. Raidal and U. Sarkar, \pha{0012101};
E. Ma and M. Raidal, \pha{0012366};
M.B. Tully and G.C. Joshi, Phys.\ Rev. {\bf D64} (2001) 011301 \phb{0011172};
Z. Tavartkiladze, \pha{0105281}.

\bibitem{Bar2} 
G. Barenboim {\it et. al.},
Nucl. Phys. {\bf B511} (1998) 577 {[{\tt hep-ph/9611347}]}.

\bibitem{Bar3} 
G. Barenboim, J. Bernabeu and M. Raidal,
\prd{55} (1997) 4213 \\ {[{\tt hep-ph/9702337}]}.

\bibitem{Bar4} 
G. Barenboim, J. Bernabeu, J. Matias and M. Raidal,
Phys. Rev. {\bf D60} (1999) 016003 \\ {[{\tt hep-ph/9901265}]}. 


\bibitem{Ball2} 
P. Ball and R. Fleischer,
\plb{475} (2000) 111 {[{\tt hep-ph/9912319}]}.


\bibitem{BP2}S. Bergmann and G. Perez, \pha{0103299}, 
Phys.\ Rev. {\bf D}, in press.



\bibitem{BaBar}B. Aubert {\it et al.}, 
BaBar collaboration, \exa{0107013}.

\bibitem{Belle}K. Abe {\it et al.},
BELLE collaboration, \exa{0107061}.


\bibitem{Beneke}M. Beneke, Plenary Talk, the EPS Int. Conference on
High Energy Physics, 
July 14 (2001-Budapest). 



\end{thebibliography}
\end{document}